\documentclass[12pt]{article}

\usepackage{amsfonts}
\usepackage{amssymb}
\usepackage{amsthm}
\usepackage[fleqn]{amsmath}
\usepackage[mathscr]{eucal}

\newtheorem{theorem}{Theorem}

\newcommand{\refeq}[1]{(\ref{#1})}

\newcommand{\pdv}{\partial_\vV}

\newcommand{\pdvk}{\partial_{\vV_k}}
\newcommand{\pdvl}{\partial_{\vV_l}}
\newcommand{\pdvuno}{\partial_{\vV_1}}
\newcommand{\pdvdue}{\partial_{\vV_2}}
\newcommand{\dd}{\mathrm{d}}

\newcommand{\pdt}{\partial_t}

\newcommand{\pdw}{\partial_\wV}

\newcommand{\vect}[1] {\boldsymbol{{ #1}} }
\newcommand{\eV}{{\vect{e}}}           
\newcommand{\pV}{{\vect{p}}}           
\newcommand{\uV}{{\vect{u}}}           
\newcommand{\vV}{{\vect{v}}}           
\newcommand{\wV}{{\vect{w}}}           
\newcommand{\eVN}{{\vect{E}}}          
\newcommand{\mVN}{{\vect{M}}}          
\newcommand{\vVN}{{\vect{V}}}          
\newcommand{\uVN}{{\vect{U}}}          
\newcommand{\wVN}{{\vect{W}}}          
\newcommand{\vVk}{{\vect{v}_k}}            
\newcommand{\alphaV}{{\vect{\alpha}}}           
\newcommand{\gammaV}{{\vect{\gamma}}}           

\newcommand{\id}{\mathbf{I}}

\newcommand{\abs}[1]{\left| #1 \right|}
\newcommand{\norm}[1]{\left\| #1 \right\| }

\newcommand{\vareps}{\varepsilon}

\newcommand{\defeg}{\stackrel{\textrm{\tiny def}}{=}}

\newcommand{\Bset}{\mathbb{B}}

\newcommand{\Dset}{\mathbb{D}}

\newcommand{\Iset}{\mathbb{I}}
\newcommand{\Lset}{\mathbb{L}}
\newcommand{\Mset}{\mathbb{M}}
\newcommand{\Nset}{\mathbb{N}}
\newcommand{\Rset}{\mathbb{R}}
\newcommand{\Sset}{\mathbb{S}}

\newcommand{\Prm}{\boldsymbol{P}}

\newcommand{\Asp}{\mathfrak{A}}
\newcommand{\Csp}{\mathfrak{C}}
\newcommand{\Hsp}{\mathfrak{H}}
\newcommand{\Gsp}{\mathfrak{G}}

\newcommand{\Lsp}{\mathfrak{L}}
\newcommand{\Nsp}{\mathfrak{N}}

\newcommand{\Wsp}{\mathfrak{W}}

\newcommand{\LCn}{{\cal L}^{(n)}}
\newcommand{\LCN}{{\cal L}^{(N)}}
\newcommand{\LCkl}{{\cal L}_{\vV_k,\vV_l}}

\newcommand{\LCvw}{{\cal L}_{\vV,\wV}}

\begin{document}

\title{On the master equation approach to\\ kinetic theory:
       linear and nonlinear\\ Fokker--Planck equations}
\vspace{-0.3cm}
\author{
\normalsize Michael Kiessling\\[-0.1cm]
\normalsize Department of Mathematics, Rutgers University\\[-0.1cm]
\normalsize Piscataway NJ 08854, USA\\[0.3cm]
\normalsize Carlo Lancellotti\\[-0.1cm]
\normalsize Department of Mathematics, City University of New York-CSI\\[-0.1cm]
\normalsize Staten Island NY 10314, USA
}
\vspace{-0.3cm}
\date{}
\maketitle
\vspace{-0.6cm}

\begin{abstract}
We discuss the relationship between kinetic equations of the Fokker-Planck type 
(two linear and one non-linear) and the Kolmogorov (a.k.a. master) equations 
of certain $N$-body diffusion processes, in the context of Kac's propagation-of-chaos limit.
The linear Fokker-Planck equations are well-known, 
but here they are derived as a limit $N\to\infty$ 
of a simple linear diffusion equation on $3N-C$-dimensional $N$-velocity spheres of 
radius $\propto\sqrt{N}$ (where $C=1$ or $4$ depending on whether the
system conserves energy only or energy and momentum).
 In this case,  a spectral gap separating the zero eigenvalue 
from the positive spectrum of the Laplacian remains as $N\to\infty$,
so that the exponential approach to equilibrium of the master evolution 
is passed on to the limiting Fokker-Planck evolution in $\Rset^3$.
 The non-linear Fokker-Planck equation is known as Landau's equation in the plasma 
physics literature. 
 Its $N$-particle master equation, originally introduced (in the 1950s) 
by Balescu and Prigogine (BP), is studied here on the ${3N-4}$-dimensional 
$N$-velocity sphere. 
  It is shown that the BP master equation represents a superposition of diffusion 
processes on certain two-dimensional sub-manifolds of $\Rset^{3N}$ 
determined by the conservation laws for two-particle collisions. 
 The initial value problem for the BP master equation is proved to be well-posed 
and its solutions are shown to decay exponentially fast to equilibrium. 
  However, the first non-zero eigenvalue of the BP operator is shown to 
vanish in the limit $N\to\infty$. 
  This indicates that the exponentially fast approach to equilibrium  
may not be passed from the finite-$N$ master equation on to Landau's nonlinear kinetic equation.
\end{abstract}
\vfill

\hrule
\smallskip
\noindent
{\small
\copyright 2004 The authors. This paper may be reproduced for noncommercial purposes.
}

\newpage


\setlength{\baselineskip}{20pt}
\section{Introduction}

 Kinetic equations play a crucial role in the transport theory of gases and plasmas,
in particular for studying the approach to equilibrium.
 Apart from rigorous mathematical studies of their solvability properties and the
classification and description of their solutions, it is essential to establish 
their {\it validity}. 
 The validation of a kinetic equation consists in its derivation from some 
deeper, deterministic microscopic model, for instance from the classical Newtonian 
dynamics of an isolated system of many interacting (point) particles.
 Clearly, a complete validation automatically involves existence and uniqueness 
results for the kinetic equation that is being validated, and it also
involves proving some version of the second law of thermodynamics.
 A priori knowledge of existence and uniqueness of solutions to 
the kinetic equation can aid the proof of its validity, while in the absence of 
such a priori knowledge the successful validation would yield existence and
uniqueness for the kinetic equation as a corollary. 
 Unfortunately, validation has turned out to be a very difficult problem.
 Even the probably most re-known and most studied of the kinetic equations, 
namely Boltzmann's equation, has been validated only in a few ``simple'' situations 
\cite{Lanf,SpoBOOK,CIPbook}.  

 Meanwhile, inspired by the pioneering work of Kac \cite{Kac} a 
large body of literature has accumulated in which the deterministic N-body dynamics is replaced
by an interacting stochastic Markov process which preserves, in each binary interaction, 
at least particle number and energy, but preferably also momentum and angular momentum, 
and which is designed to formally lead to the same kinetic equation that one expects from 
the ``physical" N-body system in the infinitely many particles limit through a law of large 
numbers.
 Technically, the law of large numbers for the stochastic evolution of a family of
individual systems of $N$ particles is equivalent to studying Kac's propagation of 
chaos limit $N\to\infty$ for the corresponding {\it ensemble} of individual systems, 
the evolution of which is being described by the Kolmogorov equation 
(called {\it master equation} in the physics literature) 
for the selected Markov process.
 Like Liouville's equation, the Kolmogorov equation is a \emph{linear} deterministic
partial differential equation for the ensemble probability density on $6N$-dimensional 
phase space.
 Unlike Liouville's equation, the Kolmogorov equation typically defines a contraction semi-group
instead of a group as does Liouville's equation.
 Hence, the relaxation of the ensemble density to a uniform density is now built into the 
ensemble evolution, and because of Boltzmann's result that in the limit $N\to\infty$ almost 
every point of phase space corresponds to the Maxwellian velocity
distribution, the  second law of thermodynamics is a foregone conclusion.
 In this sense, the Kac approach is simpler than the original validation problem  
for kinetic equations, but it still retains some flavor of validation.
 {F}rom the perspective of validation, one could say that the Kac type 
approach goes ``half the way'' toward what one would like to prove.
 {F}rom the perspective of the mathematical analysis of the kinetic equations itself, 
the Kac type approach offers a new angle of attack to establish existence and uniqueness 
of the evolution and the relaxation to equilibrium in those cases where these results have 
not yet been obtained by direct PDE methods. 

 Yet, the question whether the information obtained for a linear master (Kolmogorov)
equation for finite $N$ (such as global existence and uniqueness of solutions, as well
as exponentially fast approach to equilibrium) carries on to the typically nonlinear 
kinetic equation which is expected to arise from the master equation in the limit $N\to\infty$,
has turned out to be more subtle than originally anticipated.
 The current state of the art of this approach for short range binary processes is presumably 
the work
  \cite{CarLoss}.

 Our primary concern in this paper is the master equation approach to
certain kinetic equations that arise in the theory of systems with long range interactions, 
such as Coulomb plasmas and Newtonian gravitating systems.  
 In particular, we discuss a master equation, originally introduced (in the 1950s) 
by Balescu and Prigogine (BP), that leads formally to the spatially homogeneous Landau 
kinetic equation \cite{Land}, which plays a fundamental role in the classical transport 
theory of Coulomb plasmas \cite{BalBOOKc,Hin}.
  Considering here only the one-component case, the Landau equation 
for the particle density function $f(\, .\, ;t):\Rset^3\to\Rset_+$ 
on velocity space at time  $t\in \Rset_+$ has the form  
\begin{equation}
\pdt{f}(\vV;t)
=
\pdv\, {\cdot}\! \int_{\Rset^3} \boldsymbol{Q_L}(\vV,\wV)
\,\cdot\, \big(\pdv -\pdw\big)\big(f(\vV;t)f(\wV;t)\big)\,\dd^3\wV,
\label{LeqB}  
\end{equation}
where $\boldsymbol{Q_L}(\vV,\wV)$ is the Landau collision
kernel 
\begin{equation}
\boldsymbol{Q}_L(\vV,\wV)
= \pdw^{\otimes 2}|\vV-\wV|
=|\vV-\wV|^{-1}\,\, \Prm_{\vV-\wV}^\perp
\end{equation}
with $\Prm_{\vV-\wV}^\perp$ the projector onto the plane
perpendicular to $\vV-\wV \in \Rset^3$. 
 We remark that with the help of the so-called Rosenbluth potentials 
\cite{RMcDJ} of $f$, the Landau equation can be recast 
as a \textit{nonlinear} Fokker-Planck equation, the form that is 
better-known to the astrophysics community. 
 Formally, the Landau equation satisfies the standard conservation laws of mass, 
momentum and energy; also the H-theorem holds.  
 Thus, it is commonly believed that at late times the solution $f$ evolves into
the Maxwellian $f_M$ associated, via the conservation laws, with 
the initial data $f_0$. 
 Estimates of the relaxation time are usually
obtained by linearization of the equations, but without estimates of
the time it takes the dynamics to reach the linear regime. 
 Despite its physical importance, the mathematically rigorous confirmation
of the expected behavior of the solutions to this equation is lacking. 
 Only very recently has the spatially homogeneous Landau equation 
attracted some attention in the PDE literature, where it has been studied
as a member of a more general family of equations with kernels
\begin{equation}
\boldsymbol{Q}(\vV,\wV)
=|\vV-\wV|^{2+\gamma}\,\, \Prm_{\vV-\wV}^\perp\qquad\qquad(\gamma>-5)
\label{Qgamma}
\end{equation}
formally associated with $1/r^{\frac{\gamma-5}{\gamma-1}}$ force laws.
 However,  for Coulomb and Newton interactions ($\gamma=-3$), these PDE methods 
have provided only weak existence results \cite{Vill1,Vill2}, 
leaving the questions of uniqueness, regularity and approach to equilibrium 
largely unanswered, except for initial conditions close to equilibrium
\cite{Guo} or locally in time \cite{ZhanA}.  

As regards the Balescu--Prigogine $N$-particle master
equation for the Landau equation, we will show that its
initial value problem is well-posed, and that its solutions
approach equilibrium exponentially fast, on the ${3N-4}$-dimensional
$N$-velocity sphere of constant mass, energy, and momentum.
 However, in the limit $N\to\infty$, with energy and momentum scaled so
that the corresponding quantities per particle are constant (the $N$-velocity
sphere has radius $\propto{\sqrt{N}}$), the first non-zero eigenvalue of the BP 
operator is shown to vanish. 
  This indicates that the exponentially fast approach to equilibrium described by the 
finite-$N$ master equation may not be passed on to Landau's nonlinear kinetic equation. 
 To resolve this issue requires a more detailed knowledge of the spectrum of the 
BP operator. 

 While our efforts have not yet revealed all the details of the BP operator 
spectrum, we have discovered that the BP master equation represents a 
superposition of diffusion 
processes on certain two-dimensional sub-manifolds of $\Rset^{3N}$ 
determined by the conservation laws for two-particle collisions. 
 This has prompted us to study in more detail the completely solvable
cases in which the Kolmogorov equation is just the linear diffusion equation 
$\partial_t F = \Delta F$
on a $3N-C$-dimensional many-velocity sphere, where $C=1$ or $4$.
 The underlying stochastic processes are the perhaps simplest single + binary processes
preserving either particle number $N$ and total energy 
$\sum_{k=1}^N\frac{1}{2} \abs{\vV_k}^2=N\vareps$
($C=1$), or particle number, total energy and total momentum  
$\sum_{k=1}^N \vV_k=N\mathbf{u}$ ($C=4$).
 The results are interesting enough to deserve being included in this paper.

 Thus, we show explicitly that in the limit $N\to\infty$ we obtain an 
essentially 
linear Fokker-Planck equation for the particle density function $f(\, .\, ;t):\Rset^3\to\Rset_+$ 
on velocity (=momentum) space at time  $t\in \Rset_+$, which for $C=4$ has the form  
\begin{equation}
\partial_t f (\vV;t)
=
\partial_{\vV}\cdot\Big(\partial_{\vV}f(\vV;t)
+\frac{3}{2\vareps_0}\big(\vV - \uV\big) f(\vV;t)\Big),
\label{FokkerPlanck}
\end{equation}
for initial data $f(\vV;0) \geq 0$ having mass per particle
\begin{equation}
\int_{\Rset^3} f(\vV;0)\dd^3\vV = 1,
\label{FokkerPlanckINITf}
\end{equation}
momentum per particle
\begin{equation}
\int_{\Rset^3} \vV f(\vV;0)\dd^3\vV = \mathbf{u},
\label{FokkerPlanckINITfv}
\end{equation}
and an energy per particle
\begin{equation}
\int_{\Rset^3} \frac{1}{2}|\vV|^2 f(\vV;0)\dd^3\vV
= 
\vareps,
\label{FokkerPlanckINITfvv}
\end{equation}
which itself is a sum of the energy per particle in the center-of-mass frame, 
$\vareps_0$, and the energy per particle of the center-of-mass motion,
$\vareps_{CM} = \frac{1}{2}|\uV|^2$; viz. $\vareps = \vareps_0 +\frac{1}{2}|\uV|^2$.
 It is easy to show that with such initial data the mass per particle
$\int_{\Rset^3} f(\vV;t)\dd^3\vV$, the momentum per particle 
$\int_{\Rset^3} \vV f(\vV;t)\dd^3\vV$, and energy per particle
$\int_{\Rset^3} \frac{1}{2}|\vV|^2 f(\vV;t)\dd^3\vV$ are conserved during the
evolution.
 Moreover, for such data the solution $f$ of \refeq{FokkerPlanck} evolves, as
$t\to\infty$, into the drifting Maxwellian $f_M$ associated, via the conservation laws, 
with the initial data $f(\vV;0)$; viz.
\begin{equation}
f_M(\vV) = \left(\frac{3}{4\pi\vareps_0}\right)^{\frac{3}{2}} \exp\left( -\frac{3|\vV -\uV|^2}{4\vareps_0} \right),
\end{equation}
and it does so exponentially fast and with monotonically increasing relative entropy
\begin{equation}
S(f|f_M) = - \int_{\Rset^3} f(\vV;t)\ln \frac{f(\vV;t)}{f_M(\vV)}\dd^3\vV,
\end{equation}
so that an $H$ theorem holds.
 The treatment without momentum conservation ($C=1$) is similar; of course, $\uV$ does 
not show in this case.

 We remark that our derivation of \refeq{FokkerPlanck} together with \refeq{FokkerPlanckINITf},
\refeq{FokkerPlanckINITfv}, and \refeq{FokkerPlanckINITfvv} from an \textit{isolated} system of $N$
particles preserving energy and momentum may not be new; however, since we could not find it in
the literature, this interpretation of \refeq{FokkerPlanck} may perhaps not be so widely known.
 Indeed, \refeq{FokkerPlanck} is usually associated with an Ornstein-Uhlenbeck process for a swarm 
of individual, independent particles, the velocities (in $\Rset^3$) of which are being thermalized 
through contact with a heat bath of temperature $T$.  In this case $2\vareps_0$ in 
\refeq{FokkerPlanck} is replaced by $3T$ and the restrictions \refeq{FokkerPlanckINITfv} 
and \refeq{FokkerPlanckINITfvv} on the initial data have to be dropped. 

 In the remainder of the paper, we first set up the master equation approach for isolated
spatially uniform systems in general.
 Then, to have a simple illustration of validation \`a la Kac, we first discuss the 
diffusion equations on the $3N-C$-dimensional velocities spheres and 
derive the linear Fokker-Planck
equation(s) in the limit $N\to\infty$. 
 Then we turn to the Balescu-Prigogine master equation and its putative $N\to\infty$ limit, 
the nonlinear Fokker-Planck equation due to Landau.

\section{Ensembles of isolated systems}

\subsection{The velocity manifolds}

 Let $\{\vVN_\alpha\}_{\alpha =1}^\infty$ denote an infinite ensemble of
identically distributed random vectors taking values in $\Rset^{3N}$.
 Each vector $\vVN =(\vV_1,...,\vV_N)\in \Rset^{3N}$ represents a possible 
micro-state of an individual system of $N$ particles with velocities 
$\vV_i=(v_{i1},v_{i2},v_{i3})\in\Rset^3$. 
 The positions of the particles are assumed to be uniformly 
distributed over either a periodic box or a container with reflecting 
boundaries and have been integrated out. 
 The micro-state is assumed to evolve in time according to some stochastic process
which conserves 
\begin{equation}
m(\vVN) = N
\qquad\qquad\quad\qquad\qquad{(\textrm{mass\ of}\ \vVN)}
\label{MAofV}
\end{equation}
and 
\begin{equation}
e(\vVN) = \frac{1}{2} \sum_{k\in\Iset_N} \abs{\vV_k}^2
\ \ \qquad\qquad{(\textrm{energy\ of}\ \vVN)},
\label{ENofV}
\end{equation}
where $\Iset_N =\{1,...,N\}$; in the periodic box also 
\begin{equation}
\pV(\vVN) = \sum_{k\in\Iset_N} \vV_k
\ \qquad\qquad{(\textrm{momentum\ of}\ \vVN)}
\label{MOofV}
\end{equation}
is preserved.
 We are only interested in ensembles of $N$ particles systems 
in which all members have the same energy, or same energy and same momentum. 
 Thus, depending on the circumstances, an ensemble consists of
vectors $\vVN_\alpha$ taking values either in the  $3N-1$-dimensional manifold 
of constant energy 
\begin{equation}
\Mset^{3N-1}_{\vareps}
=
\Big\{\vVN\; :\quad
\frac{1}{2}\sum_{k\in\Iset_N} \abs{\vV_k}^2=N\vareps,
\quad \vareps > 0 \Big\},
\end{equation}
or in the $3N-4$-dimensional manifold 
of constant energy and momentum 
\begin{equation}
\Mset^{3N-4}_{\mathbf{u},\vareps}
=
\Big\{\vVN\; :\quad\sum_{k\in\Iset_N} \vV_k=N\mathbf{u}, \quad
\frac{1}{2}\sum_{k\in\Iset_N} \abs{\vV_k}^2=N\vareps,
\quad \vareps > \frac{1}{2}|\uV|^2 \Big\}.
\end{equation}
 Each such manifold is invariant under the process which generates the
evolution of an individual, isolated  $N$ body system, which in turn
traces out a trajectory on one of these manifolds $\Mset^{3N-1}_{\vareps}$
or $\Mset^{3N-4}_{\mathbf{u},\vareps}$.

 The manifold $\Mset^{3N-1}_{\vareps}$ is identical to a $3N-1$-dimensional
sphere $\Sset^{3N-1}_{\sqrt{2N\vareps}}$ of radius $\sqrt{2N\vareps}$ 
and centered at the origin of $\Rset^{3N}$;
the manifold $\Mset^{3N-4}_{\mathbf{u},\vareps}$ is identical to a $3N-4$-dimensional
sphere of radius $\sqrt{2N\vareps_0}$ and centered at 
$\uVN = (\uV,...,\uV)$, embedded in the $3(N-1)$-dimensional 
affine linear subspace of $\Rset^{3N}$ given by $\uVN + \Lset^{3N-3}$, where
$\Lset^{3N-3} \equiv \Rset^{3N}\cap\big\{\vVN\in\Rset^{3N}\; :\quad\sum_{k\in\Iset_N} \vV_k= \mathbf{0}\big\}$.
 In the following, when we write $\Sset^{3N-4}_{\sqrt{2N\vareps_0}}$, 
we mean  $\Sset^{3N-4}_{\sqrt{2N\vareps_0}}\subset \Lset^{3N-3}$.

\subsection{Master equations}

 Any ensemble $\{\vVN_\alpha\}_{\alpha =1}^\infty$ at time $t$ is characterized by a probability 
density on either $\Mset^{3N-1}_{\vareps}$ or  $\Mset^{3N-4}_{\uV,\vareps}$, 
for simplicity denoted $F^{(N)}(\vVN;t)$.
 The time evolution of $F^{(N)}(\vVN;t)$ is determined by a master equation
on $\Lsp^2\cap \Lsp^1 (\Mset^{3N-4}_{\uV,\vareps})$
or $\Lsp^2\cap \Lsp^1 (\Mset^{3N-1}_{\vareps})$
\begin{equation}
\pdt{F^{(N)}}
=
- \LCN F^{(N)},
\label{linFPeq}
\end{equation}
where $\LCN$ is a positive semi-definite operator on 
$\Lsp^2 (\Mset^{3N-4}_{\uV,\vareps})$
or $\Lsp^2 (\Mset^{3N-1}_{\vareps})$; \refeq{linFPeq} is the
Kolmogorov equation adjoint to  the underlying stochastic process. 
In general, the operator $\LCN$ has a non-degenerate smallest eigenvalue $0$ and 
corresponding eigenspace consisting of the constant functions.
 Since all particles are of the same kind, we consider only operators $\LCN$ 
and probability densities $F^{(N)}$ which are invariant under the symmetric group 
$S_N$ applied to the $N$ components in $\Rset^3$ of $\vVN$.
Also, the density $F^{(N)}(\vVN;t)$ has to satisfy the initial condition
$\lim_{t\downarrow 0}F^{(N)}(\vVN;t) = F_0^{(N)}(\vVN)$.

%
%

\section{The diffusion master equation}

 Since it is instructive to have some explicitly solvable examples, in this 
section we consider first the case of an isolated gas in a container, then we
turn to the case of an isolated gas in a periodic box.

\subsection{Gas in a container}

\subsubsection{Finite $N$}

Taking
\begin{equation}
\LCN
=
- \Delta_{\Mset^{3N-1}_{\vareps}},
\label{LAPop}
\end{equation}
the master equation is then simply the diffusion equation on $\Mset^{3N-1}_{\vareps} = \Sset^{3N-1}_{\sqrt{2N\vareps}}$,
\begin{equation}
\pdt F^{(N)}(\vVN;t)
=
 \Delta_{\Mset^{3N-1}_{\vareps}}
\,F^{(N)}(\vVN;t).
\label{heatEQonMeps}
\end{equation}
Now, the spectrum and eigenfunctions of the Laplacian on a $D$-dimensional sphere
are well-known.  Since 
$\Delta_{\Sset^{3N-1}_{\sqrt{2N\vareps}}} = 
\frac{1}{2N\vareps}\Delta_{\Sset^{3N-1}}$
it will be enough to consider them on the unit sphere;
the results can then be adapted to $\Sset^{3N-1}_{\sqrt{2N\vareps}}$ by 
simple scaling.
 With $D=3N-1$, the spectrum  of 
$-\Delta_{\Sset^{3N-1}}$ is given by 
$\{\lambda_{\Sset^{3N-1}}^{(j)}\}_{j=0}^\infty$ 
with $\lambda_{\Sset^{3N-1}}^{(j)} = j(j + 3N -2)$, 
and the eigenspace for $\lambda_{\Sset^{3N-1}}^{(j)}$
is spanned by the restrictions to 
$\Sset^{3N-1}\subset\Rset^{3N}$ 
of the harmonic polynomials which are homogeneous of degree $j$ in $\Rset^{3N}$;
when $j>0$ this restriction to $\Sset^{3N-1} \subset\Rset^{3N}$ has 
to be non-constant.
Solutions of the diffusion master equation which are invariant
under the symmetry group $S_N$ acting on $\vVN$, however, can be expanded
entirely in terms of eigenfunctions having that same symmetry.
The simplest such eigenfunctions are the restriction to 
$\Sset^{3N-1}$ of
the polynomials of the form 
$P^{(1)}_j(\vVN) = \sum_{k\in\Iset_N} p_j(\vV_k)$
where the $p_j(\vV)$'s are harmonic polynomials
which are homogeneous of degree $j$ in $\Rset^3$
(however, the special case of $p_2(\vV) = v_1^2 +v_2^2 - 2v_3^2$ simply leads to
the constant function on  $\Sset^{3N-1}$ and, 
hence, does not lead to an element of the eigenspace of 
$\lambda_{\Sset^{3N-1}}^{(2)}$).
 In the next more complicated case the $S_N$-invariant eigenfunctions for 
$j={j_1+j_2}$ are of the form
$P^{(2)}_{j_1+j_2}(\vVN) = \sum_{k\in\Iset_N} \sum_{\ l\in\Iset_{N-1}^{(k)}}
p_{j_1}(\vV_k)p_{j_2}(\vV_l)$, restricted to 
${\Sset^{3N-1}}$; etc.

 Thus, the $S_N$-symmetric solutions of equation \refeq{heatEQonMeps}
on $\Mset^{3N-1}_{\vareps} = \Sset^{3N-1}_{\sqrt{2N\vareps}}$ are given by a generalized
Fourier series
\begin{equation}
F^{(N)}(\vVN;t) 
= 
\abs{\Sset^{3N-1}_{\sqrt{2N\vareps}}}^{-1}
+ 
\sum_{j\in\Nset} \sum_{\ell\in\Dset_j} 
F_{j,\ell}^{(N)}  
G_{j,\ell}^{(N)}(\vVN) 
\exp\left(- \textstyle{\frac{j(j + 3N -2)}{2N\vareps}}t\right)
,
\label{FNevolution}
\end{equation}
where $\{G_{j,\ell}^{(N)}(\vVN),\;\ell\in \Dset_j\}$ are the eigenfunctions of 
$-\Delta_{\Sset^{3N-1}_{\sqrt{2N\vareps}}}$ spanning the $S_N$-symmetric 
eigen-subspace for $\lambda_{\Sset^{3N-1}_{\sqrt{2N\vareps}}}^{(j)}$ for
$j\in \Nset$, with $\Dset_j \subset \Nset$ the set of indices labeling the 
degeneracy of the $j$-th eigenvalue, 
and the $F_{j,\ell}^{(N)}$ are the expansion coefficients.
 (Although the eigenfunctions are quite explicitly computable, we here refrain from listing
them; we shall only work with some simple eigenfunctions for the purpose of illustration.)
 Evidently, the ensemble probability density function on 
${\Sset^{3N-1}_{\sqrt{2N\vareps}}}$ evolves  exponentially fast into the 
uniformly spread-out probability density  $\abs{\Sset^{3N-1}_{\sqrt{2N\vareps}}}^{-1}$
which is the eigenfunction for $\lambda_{\Sset^{3N-1}_{\sqrt{2N\vareps}}}^{(0)} = 0$.

\subsubsection{The limit $N\to\infty$}

 To discuss the limit $N\to\infty$ for the time-evolution of the 
ensemble, we consider the time-evolution of the hierarchy of
$n$-velocity marginal distributions $F^{(n|N)}(\vV_1,\dots,\vV_n;t)$
with domains 
$\{(\vV_1,\dots,\vV_n):
\sum_{k=1}^n |\vV_k|^2\leq {2N\vareps}\}\subset\Rset^{3n}$, 
which obtains by integrating \refeq{FNevolution} over the available 
domains ${\Sset^{3(N-n)-1}_{\sqrt{2N\vareps -\sum_{k=1}^n|\vV_k|^2}}}$ 
of the remaining $N-n$ velocities variables, thus
\begin{eqnarray}
F^{(n|N)}(\vV_1,\dots,\vV_n;t)
&=& 
F^{(n|N)}_{\mathrm{stat}}(\vV_1,\dots,\vV_n)
\nonumber\\ 
&+&\sum_{j\in\Nset} \sum_{\ell\in\Dset_j} 
F_{j,\ell}^{(N)}  
g_{j,\ell}^{(n|N)}(\vV_1,\dots,\vV_n)
e^{- \textstyle{\frac{j(j + 3N -2)}{2N\vareps}}t},
\label{FnNevolution}
\end{eqnarray}
where $F^{(n|N)}_{\mathrm{stat}}(\vV_1,\dots,\vV_n)$ and 
$g_{j,\ell}^{(n|N)}(\vV_1,\dots,\vV_n)$ are the corresponding
integrals of $\abs{\Sset^{3N-1}_{\sqrt{2N\vareps}}}^{-1}$ and 
$G_{j,\ell}^{(N)}(\vVN)$ over 
${\Sset^{3(N-n)-1}_{\sqrt{2N\vareps -\sum_{k=1}^n|\vV_k|^2}}}$.

 The $n$-velocity marginal  
of $\abs{\Sset^{3N-1}_{\sqrt{2N\vareps}}}^{-1}$
is 
\begin{equation}
F^{(n|N)}_{\mathrm{stat}}(\vV_1,\dots,\vV_n) 
=
\frac{\abs{\Sset^{3(N-n)-1}}}{\abs{\Sset^{3N-1}}}
\Big(2N\vareps\Big)^{\frac{-3n}{2}}
\Big(1-
{\textstyle{\frac{1}{2N\vareps}\sum_{k=1}^n}}|\vV_k|^2\Big)^{\frac{3(N-n)-1}{2}}
\end{equation}
and converges in the limit $N\to\infty$ to the $n$-velocity Maxwellian 
\begin{equation}
f_M^{\otimes{n}}(\vV_1,...,\vV_n) 
= 
\prod_{k=1}^n
\left({\textstyle{\frac{3}{4\pi\vareps}}}\right)^{\frac{3}{2}} 
\exp\left( -{\textstyle{\frac{3}{4\vareps}}}|\vV_k|^2 \right)
\label{nMaxwellian}
\end{equation}
with domain $\Rset^{3n}$.
 Hence, one obtains the same equilibrium statistical mechanics
from $\abs{\Sset^{3N-1}_{\sqrt{2N\vareps}}}^{-1}$, viewed 
as a probability density on $\Mset^{3N-1}_{\vareps}$, as one 
obtains from the conventional Boltzmann--Gibbs micro-canonical 
equilibrium ensemble 
$Z^{-1}\delta(\frac{1}{2}\sum_{k\in\Iset_N}|\vVk|^2 - N\vareps)$, viewed
as a stationary probability ``density'' on $\Rset^{3N}$ for the velocities
of the perfect classical gas of $N$ particles at energy $N$.
 Note that for finite $N$, the results differ slightly.

 As for the  partially integrated $S_N$-symmetric eigenfunctions of 
$-\Delta_{\Sset^{3N-1}_{\sqrt{2N\vareps}}}$, suitably normalized they 
converge pointwise to a compatible family of $S_n$-symmetric functions 
on $\Rset^{3n}$, 
\begin{equation}
\lim_{N\to\infty} g_{j,\ell}^{(n|N)}(\vV_1,\dots,\vV_n)
= 
g_{j,\ell}^{(n)}(\vV_1,\dots,\vV_n).
\end{equation}
A detailed calculation, to be presented elsewhere, shows that the limit
functions $g_{j,\ell}^{(n)}(\vV_1,\dots,\vV_n)$ are identically zero
unless $\ell$ belongs to a certain subset
$\tilde\Dset_j\subset\Dset_j$.  If $\ell\in\tilde\Dset_j$, each
$g_{j,\ell}^{(n)}(\vV_1,\dots,\vV_n)$ turns out to be, for all $n$, one of 
the well-known \cite{Risk}
eigenfunctions of the Fokker-Planck equation in $\Rset^{3n}$
(see \refeq{nFPeq} here  below). Each eigenfunction is given by
the $n$-velocity Maxwellian \refeq{nMaxwellian} multiplied by a (symmetrized)
product of Hermite polynomials, one in each component of the 
$n$ velocities $\vV_1,\dots,\vV_n$, of total degree $j$.

 Regarding the  spectrum of $-\Delta_{\Sset^{3N-1}_{\sqrt{2N\vareps}}}$, 
it is readily seen that in the limit $N\to\infty$ we have
\begin{equation}
  \lim_{N\to\infty} 
\Bigl\{
        \lambda_{\Sset^{3N-1}_{\sqrt{2N\vareps}}}^{(j)}
\Bigr\}_{j=0}^\infty 
=
\Big\{\textstyle{
        \frac{3j}{2\vareps}}
\Big\}_{j=0}^\infty.
\label{LIMspectrum}
\end{equation}
Thus,  the whole spectrum remains discrete and, as is well known \cite{Risk},  
coincides with the spectrum of the harmonic quantum oscillator (after a 
scaling and a shift).  In particular, there is a spectral gap separating the 
origin from the rest of the spectrum.

Finally, if one chooses a sequence $\{F_{j,\ell}^{(N)}\}$ that converges to an 
appropriate limit sequence $\{F_{j,\ell}\}$, and if the initial $n$-velocity 
marginal  
in the limit $N\to\infty$ is given by
\begin{equation}
f^{(n)}(\vV_1,\dots,\vV_n;0)
= 
f_M^{\otimes{n}}(\vV_1,...,\vV_n)
+ 
\sum_{j\in\Nset} \sum_{\ell\in\tilde\Dset_j} 
F_{j,\ell} 
g_{j,\ell}^{(n)}(\vV_1,\dots,\vV_n), 
\end{equation}
then its subsequent evolution is given by 
\begin{equation}
f^{(n)}(\vV_1,\dots,\vV_n;t)
= 
f_M^{\otimes{n}}(\vV_1,...,\vV_n)
+ 
\sum_{j\in\Nset} \sum_{\ell\in\tilde\Dset_j} 
F_{j,\ell} 
g_{j,\ell}^{(n)}(\vV_1,\dots,\vV_n)
e^{- \textstyle{\frac{3j}{2\vareps}}t}
\end{equation}
and describes an exponentially fast approach to equilibrium in the infinite system.
 
 Coming now to the evolution equations for the $f^{(n)}(\vV_1,...,\vV_n;t)$, 
we here only state the final result, which is a special case of the more general
one presented in the next subsection.
 In the limit $N\to\infty$, the $n$-th marginal evolution equation for the 
diffusion master equation on 
$\Mset^{3N-1}_{\vareps}=\Sset^{3N-1}_{\sqrt{2N\vareps}}$, eq. \refeq{heatEQonMeps},
becomes the (essentially linear) Fokker-Planck equation in $\Rset^{3n}$,
\begin{equation}
\pdt f^{(n)}
=
\sum_{i\in\Iset_n}
\frac{\partial}{\partial\vV_i}\cdot\Big(\frac{\partial f^{(n)}}
{\partial\vV_i}+\frac{3}{2\vareps}\vV_i\,f^{(n)}\Big);
\label{nFPeq}
\end{equation}
in particular, for $n=1$ we recover eq.\ (\ref{FokkerPlanck}) for
$f\equiv f^{(1)}$, with $\uV=\mathbf{0}$ and $\vareps_0 = \vareps$, 
and together with \refeq{FokkerPlanckINITf} and \refeq{FokkerPlanckINITfvv}.
 The momentum constraint \refeq{FokkerPlanckINITfv} on the initial data is here
immaterial. 
 
  It thus appears that we have derived eq.\ (\ref{FokkerPlanck}) as a kinetic equation.
  Appearances are, however, misleading.
  At this point (\ref{FokkerPlanck}) does not yet have the status of  a kinetic
equation; notice that Kac's concept of propagation of chaos has not entered the picture!
  In fact, $f^{(n)}$ in \refeq{nFPeq} may still in general be a convex linear
ensemble superposition of extremal states, which are products of
$n$ one-particle functions representing the velocity density function of
an actual member of the infinite ensemble.  
 In other words, \refeq{nFPeq} for $n=1,2,\dots$
defines a ``Fokker-Planck hierarchy" for a general statistical superpositions of initial
conditions, analogous to the well-known Boltzmann hierarchy that arises in the kinetic
theory of dilute gases \cite{SpoBOOK,CIPbook}.  
 In this simple case, however, the  $n$-th 
linear equation in the hierarchy \refeq{nFPeq} is decoupled from the equation
for the $n+1$-th marginal.  
  The upshot is that the first equation of the hierarchy is decoupled from $f^{(2)}$ and
therefore already a closed equation for $f^{(1)}$.
 Since it is {\it essentially linear} (we say ``essentially,'' for the
parameter $\vareps$ is coupled with the initial data), a linear superposition of
different solutions (corresponding to a statistical ensemble of initial data 
for $f$ with same energy) 
is again a solution. 
 Hence, if initially 
\begin{equation}
f^{(n)}(\vV_1,\dots,\vV_n;0)
= 
\langle f_0^{\otimes{n}}(\vV_1,...,\vV_n)\rangle,
\label{HWinitially}
\end{equation}
where $\langle\,.\,\rangle$ is the Hewitt--Savage \cite{HewSav}
ensemble decomposition measure 
on the space of initial velocity density functions of {\it individual}
physical systems, where each $f_0(\vV)\geq 0$ satisfies \refeq{FokkerPlanckINITf} and
\refeq{FokkerPlanckINITfvv}, then at later times
\begin{equation}
f^{(n)}(\vV_1,\dots,\vV_n;t)
= 
\langle f^{\otimes{n}}(\vV_1,...,\vV_n;t)\rangle
\end{equation}
where $f(\vV;t)$ solves eq.\ (\ref{FokkerPlanck}) 
with $\uV=\mathbf{0}$ and $\vareps_0 = \vareps$, 
and together with \refeq{FokkerPlanckINITf} and \refeq{FokkerPlanckINITfvv}.
 Note that the Hewitt-Savage measure is of course invariant under the evolution.
 This finally establishes the status of eq.\ (\ref{FokkerPlanck}) 
(together with its constraints) as a kinetic equation valid for (almost)
every individual member of the limiting ensemble. 

 We remark that a product structure for $f^{(n)}(\vV_1,\dots,\vV_n;0)$
imposes interesting relations on the expansion coefficients, 
but we have no space to enter their discussion here.

\subsection{Gas in a periodic box}

 A periodic box is physically unrealistic, but it provides a simple
example of a situation in which momentum conservation has to be taken 
into account, too.
 In large parts the discussion of the previous subsection carries over 
to this situation.

\subsubsection{Finite $N$}

The evolution of the ensemble of finite $N$ systems is now described by 
the diffusion equation on $\Mset^{3N-4}_{\mathbf{u},\vareps}$,
\begin{equation}
\pdt F^{(N)}(\vVN;t)
=
\Delta_{\Mset^{3N-4}_{\mathbf{u},\vareps}}\,F^{(N)}(\vVN;t).
\label{heat0}
\end{equation}
 The spectrum of the Laplacian is now given by 
$\{\lambda_{\Sset^{3N-4}_{\sqrt{2N\vareps_0}}}^{(j)}\}_{j=0}^\infty$ 
with $\lambda_{\Sset^{3N-4}_{\sqrt{2N\vareps_0}}}^{(j)} = j(j + 3N -5)$, 
and the eigenspace for $\lambda_{\Sset^{3N-4}_{\sqrt{2N\vareps_0}}}^{(j)}$ is 
spanned by the restrictions to $\Sset^{3N-4}_{\sqrt{2N\vareps_0}}\subset
\Lset^{3N-3}$ of the harmonic polynomials which are homogeneous of 
degree $j$ in $\Lset^{3N-3}$; when $j>0$ this restriction to 
$\Sset^{3N-4}_{\sqrt{2N\vareps_0}}\subset\Lset^{3N-3}$ has to be 
non-constant.
 The computation of the eigenfunctions is again straightforward, but 
the embedding of $\Sset^{3N-4}_{\sqrt{2N\vareps_0}}\subset \Lset^{3N-3}$ 
causes a minor inconvenience because  a rotation of all velocity
variables is involved to get back to the physical velocity variables. 
 We shall skip the details here, which will be presented elsewhere, and now
turn directly to the derivation of the hierarchy of
linear evolution equations for the marginal densities which obtains from
eq.\ (\ref{heat0}).

It is convenient to express the Laplacian on the right-hand side in  eq.\ (\ref{heat0})
in terms of the projection operator from $\Rset^{3N}$ to the tangent space to the
embedded manifold  $\Mset^{3N-4}_{\mathbf{u},\vareps}$.  The relevant formula is
discussed in the Appendix, eq.\ (\ref{LapBel2}).  In order to apply
eq.\ (\ref{LapBel2}) to eq.\ (\ref{heat0}),
we introduce an orthogonal basis for the orthogonal complement 
in $\Rset^{3N}$ of the tangent space to $\Mset^{3N-4}_{\mathbf{u},\vareps}$ 
at $\vVN\in\Mset^{3N-4}_{\mathbf{u},\vareps}$. 
If $\uV =\mathbf{0}$, such a basis is simply provided  by 
the set of vectors $\{\vVN, \eVN_1, \eVN_2, \eVN_3\}$, with 
$\eVN_\sigma=(\eV_{\sigma},\dots,\eV_{\sigma})$, where the $\eV_{\sigma}$, $\sigma=1,2,3$, are
the standard unit vectors in $\Rset^3$. If $\uV \neq \mathbf{0}$, the $\eVN_\sigma$ are mutually 
orthogonal but not orthogonal to 
$\vVN\in \Mset^{3N-4}_{\mathbf{u},\vareps}$; however
they are orthogonal to $\wVN\equiv\vVN-\uVN$.  Indeed,
\begin{equation}
 \vVN - \uVN 
=
\biggl(
\id_{3N} -\frac{1}{N}\sum_{\sigma=1}^3 \eVN_\sigma\otimes \eVN_\sigma
\biggr)
\cdot\vVN.
\end{equation}
 The vectors $\{\wVN, \eVN_1, \eVN_2, \eVN_3\}$ form the desired 
orthogonal basis 
for the orthogonal complement in $\Rset^{3N}$ of the tangent space to 
$\Mset^{3N-4}_{\mathbf{u},\vareps}$ at $\vVN\in\Mset^{3N-4}_{\mathbf{u},\vareps}$.
  The magnitudes of the vectors $\eVN_\sigma$ and
$\wVN$ are $\abs{\eVN_\sigma}=\sqrt{N}$ and $\abs{\wVN}=\sqrt{2N\vareps_0}$, 
respectively, where we recall that $\vareps_0 = \vareps -\frac{1}{2}\abs{\mathbf{u}}^2$.  
Then, eq.\ (\ref{heat0}) becomes
\begin{equation}
\pdt F^{(N)}
=
\frac{\partial}{\partial\vVN}\cdot \left[\left(\id_{3N}-
\frac{1}{N} \sum_{\sigma=1}^3\eVN_\sigma\otimes\eVN_\sigma-\frac{1}{2N\vareps_0}
\wVN\otimes\wVN\right)\frac{\partial F^{(N)}}{\partial\vVN}\right].
\label{heat1}
\end{equation}
 In order to obtain an equation for the $n$-th marginal of $F^{(N)}(\vVN;t)$, 
which will be denoted by $F^{(n|N)}(\vV_1,\dots,\vV_n;t)$, we integrate eq.\ (\ref{heat1}) over 
$(\vV_{n+1},\dots,\vV_N)\in \Rset^{3N-3n}$. 
 Clearly,
\begin{equation}
\int d\vV_{n+1}\dots\, d\vV_N\, \frac{\partial}{\partial\vVN}\cdot
\left(\frac{\partial F^{(N)}}{\partial\vVN}\right)
=
\sum_{i\in\Iset_n}
\frac{\partial}{\partial\vV_i}\cdot\frac{\partial F^{(n|N)}}{\partial\vV_i}
\end{equation}
Also,
\begin{equation}\int d\vV_{n+1}\dots\, d\vV_N\,  \frac{\partial}{\partial\vVN}\cdot
\left(\sum_{\sigma=1}^3\eVN_\sigma\otimes\eVN_\sigma\,
\frac{\partial F^{(N)}}{\partial\vVN}\right)=
\sum_{k=1}^3\sum_{i,j\in\Iset_n}
\frac{\partial^2 F^{(n|N)}}{\partial v_{ik}\partial v_{jk}}\end{equation}
Finally,
\[\int d\vV_{n+1}\dots\, d\vV_N\,\frac{\partial}{\partial\vVN}\cdot
\left(\wVN\otimes\wVN\,\frac{\partial F^{(N)}}{\partial\vVN}\right)= \]
\[\sum_{i\in\Iset_n}
\frac{\partial}{\partial\vV_i}\cdot\left(\wV_i\int d\vV_{n+1}\dots\, d\vV_N\,
\sum_{j\in\Iset_N}\wV_j\cdot\frac{\partial F^{(N)}}{\partial \vV_j}\right)=\]
\[
\sum_{i,j\in\Iset_n}\frac{\partial}{\partial\vV_i}\cdot
\left(\wV_i\,\wV_j\cdot
\frac{\partial F^{(n|N)}}{\partial\vV_j}\right)+\]
\[(N-n)\sum_{i\in\Iset_n}\frac{\partial}{\partial\vV_i}\cdot\left(\wV_i
\int d\vV_{n+1}\dots\, d\vV_N\,\wV_N\cdot\frac{\partial F^{(N)}}
{\partial\vV_N}\right)=\]
\begin{equation}
\sum_{i,j\in\Iset_n}\frac{\partial}{\partial\vV_i}\cdot
\left(\wV_i\,\wV_j\cdot
\frac{\partial F^{(n|N)}}{\partial\vV_j}\right)
-3(N-n)\sum_{i\in\Iset_n}\frac{\partial}{\partial\vV_i}\cdot\left(\wV_i
F^{(n|N)}\right)\end{equation}
where $\wV_i\equiv\vV_i-\mathbf{u}$ and the permutation symmetry of 
$F^{(N)}$ was used. 
 Putting
everything together, if $F^{(N)}$ satisfies the diffusion equation on 
$\Mset^{3N-4}_{\mathbf{u},\vareps}$ it follows that the $n$-th marginal  
$F^{(n|N)}$ satisfies
\begin{eqnarray}
\pdt F^{(n|N)}&=&
\sum_{i\in\Iset_n}
\frac{\partial}{\partial\vV_i}\cdot\frac{\partial F^{(n|N)}}{\partial\vV_i}-
\frac{1}{N}\sum_{k=1}^3\sum_{i,j\in\Iset_n}
\frac{\partial^2 F^{(n|N)}}{\partial v_{ik}\partial v_{jk}}
\nonumber\\
&&
-\frac{1}{2N\vareps_0}
\sum_{i,j\in\Iset_n}\frac{\partial}{\partial\vV_i}\cdot
\left((\vV_i-\mathbf{u})\,(\vV_j-\mathbf{u})\cdot 
\frac{\partial F^{(n|N)}}{\partial\vV_j}\right)
\nonumber\\
&&
+\frac{3(N-n)}{2\vareps_0N}\sum_{i\in\Iset_n}
\frac{\partial}{\partial\vV_i}\cdot\Big((\vV_i-\mathbf{u}) F^{(n|N)}\Big)
\label{nDIFFhierarchyEQ}
\end{eqnarray}

\subsubsection{The limit $N\to\infty$}

 The  spectrum of $-\Delta_{\Sset^{3N-4}_{\sqrt{2N\vareps_0}}}$
in the limit $N\to\infty$ is given by
\begin{equation}
  \lim_{N\to\infty} 
\Bigl\{
        \lambda_{\Sset^{3N-4}_{\sqrt{2N\vareps_0}}}^{(j)}
\Bigr\}_{j=0}^\infty 
=
\Big\{\textstyle{
        \frac{3j}{2\vareps_0}}
\Big\}_{j=0}^\infty,
\label{LIMspectrumNEw}
\end{equation}
which up to the replacement of $\vareps$ by $\vareps_0$ 
agrees with \refeq{LIMspectrum}.
 Thus,  the whole spectrum is once again discrete and, 
in particular, there is a spectral gap separating the 
origin from the rest of the spectrum.

 As a result, once again the velocity densities $f^{(n)}$ 
approach equilibrium  exponentially fast.
 The evolution equation for  $f^{(n)}$ which obtains in the limit $N\to\infty$
from \refeq{nDIFFhierarchyEQ} is the (essentially linear) Fokker-Planck
equation in $\Rset^{3n}$,
\begin{equation}
\pdt f^{(n)}
=
\sum_{i\in\Iset_n}
\frac{\partial}{\partial\vV_i}\cdot\Big(\frac{\partial f^{(n)}}
{\partial\vV_i}+\frac{3}{2\vareps_0}(\vV_i-\mathbf{u})\,f^{(n)}\Big)
\label{nDIFFhierarchyEQlim}
\end{equation}
 In particular, for $n=1$ we recover eq.\ (\ref{FokkerPlanck}) with 
$f\equiv f^{(1)}$, together with the constraints on the initial data
\refeq{FokkerPlanckINITf}, \refeq{FokkerPlanckINITfv}, 
and \refeq{FokkerPlanckINITfvv}.
 The last step to establish the status of (\ref{FokkerPlanck}) 
(together with its constraints) as a kinetic equation involves once
again the Hewitt--Savage decomposition, which implements Kac's
concept of propagation of chaos for \refeq{nDIFFhierarchyEQlim}.

\section{The Balescu-Prigogine master equation}

 After having established that the linear Fokker-Planck equation 
(\ref{FokkerPlanck}) together with the constraints on the initial data
can be derived as a kinetic equation from eq.\ (\ref{heat0}), the diffusion 
equation on the energy-momentum foliation of $\Rset^{3N}$, we now study 
the more complicated diffusion process on the foliation 
$\Mset^{3N-4}_{\mathbf{u},\vareps}$ associated with the (nonlinear) Landau equation, 
eq.\ (\ref{LeqB}).
 At least at the formal level, the Kolmogorov equation for the diffusion 
process in question is given by the \emph{Balescu-Prigogine master equation} 
\cite{PriBal2} 
for the time evolution of $F^{(N)}$, which can be written once again as
\begin{equation}
\pdt{F^{(N)}}
=
- \LCN F^{(N)},
\label{BPode}
\end{equation}
where now
\begin{equation}
\LCN
=
\frac{1}{N-1}
\sum_{k\in\Iset_N}
\sum_{\ l\in\Iset_{N-1}^{(k)}}\LCkl
\label{linBPop}
\end{equation}
with
\begin{equation}
\LCvw
=
-\frac{1}{2}\big(\pdv - \pdw\big) {\cdot}
\Big(|\vV-\wV|^{-1}\,\Prm_{\vV-\wV}^\perp\,\cdot\,\big(\pdv - \pdw\big)\,\Big).
\label{Lop}
\end{equation}
and $\Iset_{N-1}^{(k)} =\Iset_N\backslash\{k\}$;
the density $F^{(N)}$ has to satisfy the initial condition
$\lim_{t\downarrow 0}F^{(N)}(\vVN;t) = F_0^{(N)}(\vVN)$.
A Fourier-transformed version of \refeq{BPode} was constructed 
in~\cite{PriBal1,PriBal2}, but it does not seem to have received
much attention ever since.  
Balescu and Prigogine already pointed out
that Landau's equation \refeq{LeqB} can be extracted from \refeq{BPode} 
by contraction onto the first marginal of $F^{(N)}(\vVN;t)$.
        The formal argument runs as follows:
Eq.\ \refeq{BPode} is equivalent to a hierarchy of
evolution equations for all the marginals of $F^{(N)}(\vVN;t)$, which
are denoted by $F^{(n|N)}(\vV_1,\dots,\vV_n;t)$, with
$n=1,...,N$.  
Of course, $F^{(N|N)}\equiv F^{(N)}$.
The time evolution of $F^{(n|N)}(\vV_1,\dots,\vV_n;t)$ is given by the corresponding contraction of
the Balescu-Prigogine master equation onto $n$ variables,
\begin{equation}
\pdt{F^{(n|N)}}
=
- \frac{n-1}{N-1} \LCn F^{(n|N)}
- \frac{N-n}{N-1} \sum_{k=1}^n\int {{\cal L}_{\vV_k,\vV_{n+1}}} F^{(n+1|N)}
\dd^3\vV_{n+1}
\label{BPmargODE}
\end{equation}
Introducing the short-hands $\vV$ for $\vV_1$ and $\wV$ for $\vV_2$,
the evolution equation for the first marginal takes the form
\begin{equation}
\pdt{F^{(1|N)}(\vV;t)}
=
\pdv\, {\cdot}\! \int_{\Rset^3} |\vV-\wV|^{-1}
\, \Prm_{\vV-\wV}^\perp
\,\cdot\, \big(\pdv -\pdw\big)  F^{(2|N)}(\vV,\wV;t)\, \dd^3\wV.
\label{BBGKYeqONE}
\end{equation}
        Clearly, if we had
$F^{(2|N)}(\vV,\wV;t) = F^{(1|N)}(\vV;t)F^{(1|N)}(\wV;t)$, then
\refeq{BBGKYeqONE} would be a closed equation for $F^{(1|N)}$.
        However, since \refeq{BPode} with \refeq{linBPop} and \refeq{Lop}
does not preserve a product structure of $F^{(N)}$ for any finite $N$,
$F^{(2|N)}$ cannot remain a product of the first marginals even if
that is the case initially.
 It was Kac's insight that if one lets $N\to\infty$ and assumes that
$\lim_{N\uparrow\infty} {F^{(N)}_0(\vVN)}
=
\bigotimes_{k=1}^\infty f_0(\vV_k)$,
then one ought to be able to show that the product structure persists in time, viz.
$\lim_{N\uparrow\infty} {F^{(N)}(\vVN;t)}
=
\bigotimes_{k=1}^\infty f(\vV_k;t)$ for all $t>0$,
for which Kac coined the phrase ``the propagation of molecular chaos.''
Kac proved propagation of chaos for his caricature of the Maxwellian gas,
and Balescu and Prigogine surmised that propagation of chaos will hold
also for their master equation when $N\to \infty$, in which limit  equation
\refeq{BBGKYeqONE} then becomes
the Landau equation \refeq{LeqB}. We remark that, for initial conditions that 
do not necessarily factorize, in the $N\to\infty$ limit 
eq.\ (\ref{BPmargODE}) leads (formally) to the
\textit{Landau hierarchy}
\begin{equation}
\pdt{f^{(n)}}(\vV_1,\dots,\vV_n;t)
=
- \sum_{k=1}^n\int_{\Rset^3} {\cal L}_{\vV_k,\vV_{n+1}} f^{(n+1)}(\vV_1,\dots,\vV_{n+1};t)\,\dd^3\vV_{n+1}
\label{LANDAUnODE}
\end{equation}

 A rigorous justification of propagation of chaos for the Balescu-Prigogine
equation is an interesting  open problem, on which we recently made some
progress.
 Here we report on our results:
\begin{enumerate}
\item the Balescu-Prigogine diffusion operator has an interesting geometric 
interpretation: it is a weighted average of the Laplacians associated with 
a certain family of sub-manifolds of $\Mset^{3N-4}_{\mathbf{u},\vareps}$.  
These sub-manifolds are determined by the conservation laws for binary 
collisions between particles, as will be made clear below.
\item for each finite $N$ the Balescu-Prigogine master equation is well-posed 
in $\Lsp^2\cap\Lsp^1\big(\Mset^{3N-4}_{\mathbf{u},\vareps}\big)$ 
and displays exponential decay to equilibrium.
\item as $N\to\infty$, the first non-zero eigenvalue converges to zero. 
\end{enumerate}

 Our results show on the one hand that the finite $N$ Balescu-Prigogine equation is more similar to 
the diffusion equation than meets the eye, yet on the other hand, 
the limit $N\to\infty$ is markedly different.
 In particular, our results suggest that the spectral gap vanishes.

\subsection{Geometric Aspects of the BP Master Equation}

Clearly eq.\ (\ref{BPode}) is a linear parabolic partial differential equation
in $3N$ variables, and it is easily verified that $\LCN$,
like $\Delta_{\Mset^{3N-4}_{\mathbf{u},\vareps}}$,
annihilates the constant function $m(\vVN)$, 
the linear $3$-vector polynomial $\pV(\vVN)$, 
and the quadratic polynomial $e(\vVN)$, 
ensuring the conservation of mass, energy and momentum.  
 Hence, also in this case the evolution of the ensemble factors into independent evolutions
on the invariant manifolds $\Mset^{3N-4}_{\mathbf{u},\vareps}$.

Now, for $1\le l<k\leq N$  consider the family of $N(N-1)/2$ two-dimensional 
manifolds
\begin{equation}
\Bset^2_{kl}
=
\Big\{\vVN\ :\ \vV_k+\vV_l=\alphaV_{kl}\quad
\abs{\vV_k-\vV_l}^2=\beta_{kl}^2
\quad \vV_i=\gammaV_{kl}^{(i)},\ i\neq k,l \Big\}
\end{equation}
where $\alphaV_{kl}$, $\beta_{kl}$ and $\gammaV_{kl}^{(i)}$ are $3N-2$ 
arbitrary constants.
Clearly, each point in $\Rset^{3N}$ determines one such family of manifolds 
and each manifold corresponds to the conservation laws associated with the 
``collision" between particles $k$ and $l$.  In fact, the conservation of  
$\abs{\vV_k-\vV_l}^2$ is equivalent to the conservation 
of the two-particle energy $\abs{\vV_k}^2+\abs{\vV_l}^2$, as long as the 
two-particle momentum $\vV_k+\vV_l$ is also constant; thus, if 
$\vVN\in\Mset^{3N-4}_{\mathbf{u},\vareps}$ it follows that the
manifolds $\Bset^2_{kl}$ determined by $\vVN$ are sub-manifolds of
$\Mset^{3N-4}_{\mathbf{u},\vareps}$ for all $k,l$.
\begin{theorem}
Let $\LCvw$ be the operator in eq.\ (\ref{Lop}) and $F^{(N)}$ a
probability density on $\Mset^{3N-4}_{\mathbf{u},\vareps}$.  Then
\begin{equation}
\LCkl F^{(N)}=-\abs{\vV_k-\vV_l}^{-1}\,\Delta_{\Bset^2_{kl}}F^{(N)}
\label{BPLap}
\end{equation}
where $\Delta_{\Bset^2_{kl}}$ is the Laplace-Beltrami operator on 
$\Bset^2_{kl}$.
\end{theorem}
\begin{proof}
We first observe that the factor $\abs{\vV_k-\vV_l}^{-1}$ can be moved out
of the differential operator in eq.\ (\ref{Lop}), since 
$\big(\pdvl - \pdvk\big)\abs{\vV_k-\vV_l}^{-1}$ yields 
terms parallel to $\vV_k-\vV_l$, which are
annihilated by the projector $\Prm_{\vV_k-\vV_l}^\perp$.  
Hence, 
\begin{equation}
\LCkl
=
-\frac{1}{2}
|\vV_k-\vV_l|^{-1}\,\big(\pdvl - \pdvk\big) {\cdot}
\Big(\Prm_{\vV_k-\vV_l}^\perp\,\cdot\,
\big(\pdvk - \pdvl\big)\,\Big).
\label{Lop2}
\end{equation}
Next, we calculate $\Delta_{\Bset^2_{kl}}$.  At any point $\vVN\in\Rset^{3N}$ 
the manifold
$\Bset^2_{kl}$ has the $3N-2$ normal unit vectors $\mVN_\sigma$, $\eVN$ and
$\wVN_{i,\sigma}$ where 
\begin{equation}
\mVN_\sigma\equiv\frac{1}{\sqrt{2}}\left(
\begin{array}{c}
\mathbf{0}\\ \vdots\\ \mathbf{0}\\ \eV_\sigma\\ \mathbf{0}\\\vdots\\
\mathbf{0}\\ \eV_\sigma\\ \mathbf{0}\\ \vdots\\ \mathbf{0}
\end{array}
\right)\qquad
\eVN\equiv\frac{1}{\sqrt{2}|\vV_k-\vV_l|}\left(
\begin{array}{c}
\mathbf{0}\\ \vdots\\ \mathbf{0}\\\vV_k-\vV_l \\ \mathbf{0}\\\vdots\\
\mathbf{0}\\ -\vV_k+\vV_l\\ \mathbf{0}\\ \vdots\\ \mathbf{0}
\end{array}
\right)\quad
\begin{array}{c}
(k\ \mathrm{entry})\\
\\
\\
\\[0.2cm]
(l\ \mathrm{entry})\\
\end{array}
\end{equation}
and the $\eV_\sigma$, $\sigma=1,2,3$, are again the standard unit vectors in 
$\Rset^3$; $\wVN_{i,\sigma}$ is the standard unit vector in $\Rset^{3N}$ with 
$\eV_\sigma$ as the $i$-th 3-block, $i\neq k,l$, and zeroes everywhere else.
Since all these vectors are mutually orthogonal, the projector on the tangent
space to $\Bset^2_{kl}$ is
\begin{equation}
\mathrm{P}_{\Bset^2_{kl}}
=
\id_{3N}-\eVN\otimes\eVN-\sum_{\sigma}\mVN_\sigma\otimes\mVN_\sigma
-\sum_{i,\sigma}\wVN_{i,\sigma}\otimes\wVN_{i,\sigma}=
\end{equation}
\vspace{-0.5cm}
\[
\frac{1}{2}
\left(
\begin{array}{ccccc}
&\vdots&&\vdots&\\
\dots&\Prm_{\vV_k-\vV_l}^\perp&\dots &-\Prm_{\vV_k-\vV_l}^\perp&\dots\\
&\vdots&&\vdots&\\
\dots&-\Prm_{\vV_k-\vV_l}^\perp&\dots &\Prm_{\vV_k-\vV_l}^\perp&\dots\\
&\vdots&&\vdots&\\
\end{array}
\right)
\begin{array}{c}
\\[-0.3cm]
k
\\
\\[0.1cm]
l\\
\end{array}
\]
\vspace{-.3cm}
\[
\hspace{2.5cm}k\hspace{2.6cm}l
\]
where all the unmarked entries are zero and, of course,
\begin{equation}
\Prm_{\vV_k-\vV_l}^\perp=\id_3-\frac{\vV_k-\vV_l}{\abs{\vV_k-\vV_l}}\otimes
\frac{\vV_k-\vV_l}{\abs{\vV_k-\vV_l}}
\end{equation}
It follows immediately from eq.\ (\ref{LapBel2}) that
\begin{equation}
\Delta_{\Bset^2_{kl}}
=
\frac{1}{2}
\big(\pdvl - \pdvk\big) {\cdot}
\Big(\Prm_{\vV_k-\vV_l}^\perp\,\cdot\,
\big(\pdvk - \pdvl\big)\,\Big)
\end{equation}
which, together with eq.\ (\ref{Lop2}) gives eq.\ (\ref{BPLap}).
\end{proof}

Hence, the Balescu-Prigogine master equation can be written as
\begin{equation}
\pdt{F^{(N)}}
=
-\frac{1}{N-1}
\sum_{k\in\Iset_N}
\sum_{\ l\in\Iset_{N-1}^{(k)}}\abs{\vV_k-\vV_l}^{-1}\,\Delta_{\Bset^2_{kl}}F^{(N)}
\label{BPeq2}
\end{equation}
It is instructive to compare eq.\ (\ref{BPeq2}), which leads formally to the
Landau equation, to the diffusion equation on $\Mset^{3N-4}_{\mathbf{u},\vareps}$,
eq.\ (\ref{heat0}), which leads to the linear Fokker-Planck equation. In the
``Landau" case the diffusion does not take place isotropically over the
manifold of constant energy and momentum, but on the collection
of the sub-manifolds $\Bset^2_{kl}$ determined by the two-particle conservation
laws.  To be precise, the Balescu-Prigogine operator in eq.\ (\ref{BPeq2})
is a sort of weighted average of the Laplacians on all the 
$\Bset^2_{kl}$, where the $kl$-Laplacian has weight (``diffusivity")
$\abs{\vV_k-\vV_l}^{-1}$.  These diffusivities are constant quantities  
on the corresponding manifolds, but they can take arbitrarily small 
values in 
certain regions of $\Rset^{3N}$ as $N$ increases.  Loosely speaking,
the ellipticity of the BP equation degenerates as $N\to\infty$. This is bound
to affect the rate of decay to equilibrium of the solutions in the same limit.

\subsection{Well-posedness of the BP master equation and decay to equilibrium}

	We now identify the evolution of $F^{(N)}(\vVN;t)$ with the motion 
of a point $\psi_t$ in the Hilbert space  
$\Lsp^2\big(\Mset^{3N-4}_{\mathbf{u},\vareps}\big)$. 
We define the Sobolev-type space $\Hsp$ as the closure of 
$\Csp^\infty\big(\Mset^{3N-4}_{\mathbf{u},\vareps}\big)$ w.r.t. 
the norm $\norm{\ .\ }_{\Hsp}$, given by
\begin{equation}
\norm{\psi}_{\Hsp}^2 
\defeg 
Q^{(N)}(\psi,\psi) + \norm{\psi}_{\Lsp^2(\Mset^{3N-4}_{\mathbf{u},\vareps})}^2,
\end{equation}
where $Q^{(N)}$ is the manifestly symmetric 
positive semi-definite quadratic form associated with
the operator $\LCN$ defined on 
$\Csp^\infty\big(\Mset^{3N-4}_{\mathbf{u},\vareps}\big)$:
\begin{equation}
Q^{(N)}(\psi,\phi)
=
\int_{\Mset^{3N-4}_{\mathbf{u},\vareps}}\psi\,\LCN \phi\,\dd\tau
\label{bilinFPopN}
\end{equation}
for $(\psi,\phi)$ in 
$\Csp^\infty\big(\Mset^{3N-4}_{\mathbf{u},\vareps}\big)^{\times 2}$.
	The form closure of $Q^{(N)}(\psi,\phi)$ in $\Hsp\times\Hsp$,
which is also denoted by $Q^{(N)}(\psi,\phi)$, defines a unique self-adjoint, 
positive semi-definite operator with dense domain $\tilde\Hsp\subset\Hsp$, the 
Friedrichs extension of $\LCN$, also denoted by $\LCN$.  We recall that
the Sobolev-type space $\Hsp$ coincides with the domain of the square root
of the Friedrichs extension of $\LCN$, i.e. the
operator $[{\LCN}]^{\frac{1}{2}}$ ; see~\cite{RS1} for general background 
material.
	Thus $\LCN$ is a densely defined, unbounded operator on
the Hilbert space $\Lsp^2\big(\Mset^{3N-4}_{\mathbf{u},\vareps}\big)$. 

	It is easily seen that the kernel space of 
$\mathrm{Ker}(\LCN)\equiv\Nsp_0$ is one-dimensional, consisting of the constant 
functions.  Hence, $\Lsp^2\big(\Mset^{3N-4}_{\mathbf{u},\vareps}\big)$ decomposes 
as $\Nsp_0 \oplus \Lsp^{2,+}$, where $\Lsp^{2,+}$
is the orthogonal complement of $\Nsp_0$ in 
$\Lsp^2\big(\Mset^{3N-4}_{\mathbf{u},\vareps}\big)$, i.e. the 
subspace of $\Lsp^2\big(\Mset^{3N-4}_{\mathbf{u},\vareps}\big)$ 
on which $\LCN$ is strictly positive.  Also, $\Hsp$ decomposes as
$\Nsp_0 \oplus\Hsp^+$; we remark that $\Hsp^+\subset \Lsp^{2,+}$ 
can be equivalently defined as the closure of 
$\{\psi\in {\Csp^\infty}\big(\Mset^{3N-4}_{\mathbf{u},\vareps}\big):
\int_{\Mset^{3N-4}_{\mathbf{u},\vareps}}\psi\dd\tau =0\}$ 
w.r.t. the norm $\norm{\ .\ }_{\Hsp^+}$, where
\begin{equation}
\norm{\psi}_{\Hsp^+}^2 \defeg Q^{(N)}(\psi,\psi).
\end{equation}
 Since $\LCN$ is self-adjoint and strictly positive on $\Hsp^+\cap\tilde\Hsp$,
it follows immediately that the operator $\LCN$ 
is the generator of the contraction semi-group $e^{-t\LCN}$ 
on $\Lsp^{2,+}$~\cite{RS2}.  This implies that $\LCN$ is also
the generator of a strongly continuous semi-group on all of 
$\Lsp^2\big(\Mset^{3N-4}_{\mathbf{u},\vareps}\big)$, also denoted by
$e^{-t\LCN}$.  Precisely, $e^{-t\LCN}$ acts isometrically on the 
invariant subspace $\Nsp_0$ and strictly contracting on its
orthogonal complement $\Lsp^{2,+}$; hence $e^{-t\LCN}$ is also
positivity preserving.  Thus, if 
$\psi_0^{(N)}\in\Lsp^2\big(\Mset^{3N-4}_{\mathbf{u},\vareps}\big)$, then 
\begin{equation}
{\psi_t^{(N)}}
\defeg
e^{-t\LCN}\psi^{(N)}_0
\label{FPeqnSOL}
\end{equation}
solves \refeq{BPode} uniquely for
Cauchy data $\lim_{t\downarrow 0}{\psi^{(N)}} = {\psi^{(N)}_0}$,
and the initial value problem for the BP master equation is well-posed.
The evolution of an initial density  $\psi^{(N)}_0 \in \Lsp^1_{+,1}
\cap\Lsp^2(\Mset^{3N-4}_{\mathbf{u},\vareps})$ 
(i.e., ${\psi^{(N)}_0} >0$ and 
$
\int_{\Mset^{3N-4}_{\mathbf{u},\vareps}}
{\psi^{(N)}_0} \dd\tau = 1
$),
actually takes place in the intersection of the positive cone 
of $\Lsp^2\big(\Mset^{3N-4}_{\mathbf{u},\vareps}\big)$ 
with the affine Hilbert space
\begin{equation}
\Asp =  \psi^{(N)}_\infty + \Lsp^{2,+},
\end{equation}
where $\psi^{(N)}_\infty \defeg \abs{\Mset^{3N-4}_{\mathbf{u},\vareps}}^{-1}$ 
is the Hilbert space vector of the (analog of the) micro-canonical equilibrium ensemble.

	The spectrum of $\LCN$ can be studied with the standard techniques 
developed for weak solutions of linear inhomogeneous PDE 
in divergence form \cite{TrudGil},
extended \cite{Heb} to operators on compact manifolds without boundary 
(here $\Mset^{3N-4}_{\mathbf{u},\vareps}$).  
Some care must be taken due to the fact that the ellipticity is \emph{not 
uniform}, since the coefficients are unbounded above; on the other hand, the 
ellipticity condition is satisfied uniformly from below.
First of all, since the bilinear form $Q^{(N)}$ is clearly positive and bounded 
on $\Hsp^+$, by the Lax-Milgram Theorem it follows that for $\lambda <0$ the 
operator $\LCN_\lambda\equiv\LCN+\lambda {\cal I}$ determines a 
bijective mapping from 
$\Hsp^+$ to $(\Hsp^+)^*$ (the dual of $\Hsp^+$).  Next, we introduce a
compact embedding $\cal E$ from $\Hsp^+$ to $(\Hsp^+)^*$ such that
$({\cal E}u)(v)=\int_{\Mset^{3N-4}_{\mathbf{u},\vareps}}uv\, d\tau$ for 
all $v\in\Hsp^+$.  
 In order to define the operator $\cal E$, we first observe that 
points on $\Mset^{3N-4}_{\mathbf{u},\vareps}$ satisfy
\begin{equation}
|\vV_k-\vV_l|^2
=
\abs{\vV_k}^2+\abs{\vV_l}^2-2\vV_k\cdot\vV_l
\leq
2(\abs{\vV_k}^2+\abs{\vV_l}^2)\leq 4N\vareps
\end{equation}
so that
\begin{equation}
Q^{(N)}(\psi,\psi)
\geq
\frac{1}{2\sqrt{2N\vareps} (N-1)}
\sum_{k\in\Iset_N}
\sum_{\ l\in\Iset_{N-1}^{(k)}}
\hat Q^{(N)}_{k,l}(\psi,\psi)
\label{ineq1}
\end{equation}
where
\begin{equation}
\hat Q^{(N)}_{k,l}(\psi,\phi)
=
\frac{1}{2}
\int_{\Mset^{3N-4}_{\mathbf{u},\vareps}}
\big(\pdvk - \pdvl\big)\psi\, {\cdot}\,
\Prm_{\vV_k-\vV_l}^\perp\cdot
\big(\pdvk - \pdvl\big)\phi\,\dd\tau.
\end{equation}
Since $\Mset^{3N-4}_{\mathbf{u},\vareps}$ is a compact manifold,
it is easy to see that there is some constant $C_N$ such that
\begin{equation}
\sum_{k\in\Iset_N} \sum_{\ l\in\Iset_{N-1}^{(k)}} \hat Q^{(N)}_{k,l}(\psi,\psi) 
\geq
C_N \int_{\Mset^{3N-4}_{\mathbf{u},\vareps}}
\left|\nabla\psi\right|^2\,\dd\tau
\label{Dirich}
\end{equation}
where $\nabla\psi$ is the covariant derivative of $\psi$ on 
$\Mset^{3N-4}_{\mathbf{u},\vareps}$
and $\left|\nabla\psi\right|^2=g^{ij}\partial_i\psi\,\partial_j\psi$, $g^{ij}$
being the metric tensor and $\partial_j\psi$ the derivative with respect
to the $j$-th coordinate.  Combined with eq.\ (\ref{ineq1}) this gives
\begin{equation}
\norm{\psi}_{\Hsp^+}^2
\geq
\frac{C_N}{2\sqrt{2N\vareps}(N-1)}
\norm{\psi}_{\dot\Wsp^{1,2}_+}^2
\label{ineq2}
\end{equation} 
where $\dot\Wsp^{1,2}_+$ is the closure of
$\{\psi\in {\Csp^\infty}\big(\Mset^{3N-4}_{\mathbf{u},\vareps}\big):
\int_{\Mset^{3N-4}_{\mathbf{u},\vareps}}\psi\dd\tau =0\}$
w.r.t. the norm
\begin{equation}
\norm{\psi}_{\dot\Wsp^{1,2}_+}^2 
\defeg 
\int_{\Mset^{3N-4}_{\mathbf{u},\vareps}}
\left|\nabla\psi\right|^2\,\dd\tau.
\end{equation}
Equation (\ref{ineq2}) implies that $\Hsp^+$ is continuously embedded in
$\dot\Wsp^{1,2}_+$, which in turn is compactly embedded in $\Lsp^{2,+}$
by the Sobolev embedding theorem and the
Rellich-Kondrasov theorem, both of which hold on a compact manifold \cite{Heb}.
Finally, $\Lsp^{2,+}$ is continuously embedded in $(\Hsp^+)^*$ (via the Riesz 
Representation Theorem), and the
compact embedding $\cal E$ of $\Hsp^+$ into $(\Hsp^+)^*$ is obtained as 
the composition $\Hsp^+\to \dot\Wsp^{1,2}_+\to \Lsp^{2,+}\to (\Hsp^+)^*$. 

Finally, one obtains the standard Fredholm alternative~\cite{TrudGil}, 
by re-writing the equation $\LCN_\lambda\psi=\sigma$ as
\begin{equation}
\psi+(\lambda-\lambda_0)\Gsp_{\lambda_0}^{(N)}{\cal E}\psi
=
\Gsp_{\lambda_0}^{(N)}\sigma
\label{Fredholm}
\end{equation}
where $\Gsp_{\lambda_0}^{(N)}$ ($\lambda_0<0$) is the (continuous) inverse of 
$\LCN_{\lambda_0}$.  Then, since $\cal E$ is compact,
$-(\lambda-\lambda_0)\Gsp_{\lambda_0}^{(N)}{\cal E}$ is also compact from 
$\Hsp^+$ to $\Hsp^+$, and the standard Riesz-Schauder theory of compact
operators in a Hilbert space leads to the conclusion that $\LCN$ has a purely
discrete spectrum and that each eigenvalue has a finite-dimensional eigenspace.
 Since the spectrum of $\LCN$ is discrete, we have a spectral gap between 
$\lambda_0^{(N)} =0$ and the smallest non-zero eigenvalue $\lambda_1^{(N)} >0$ 
of $\LCN$, and we conclude that the equilibrium ensemble is 
approached exponentially in time
\begin{equation}
\norm{\psi^{(N)}_t - \psi^{N}_\infty}_{\Lsp^2}\!
=
\norm{e^{-t\LCN}\Big(\psi^{(N)}_0 - \psi^{N}_\infty\Big)}_{\Lsp^2}\!
\leq 
e^{-t\lambda_1^{(N)}}\norm{\psi^{(N)}_0 -\psi^{N}_\infty}_{\Lsp^2}\!.
\end{equation}

 \subsection{The limit $N\to\infty$}

 The question whether the spectral gap remains finite in the
limit $N\to\infty$, which was recently answered affirmatively 
in the context of  the Kac model and other models related to 
the Boltzmann equation \cite{CarLoss}, and also by us in the previous
section for the diffusion master equation, has presumably a negative
answer for the BP master equation, at least for the Coulomb case
studied here. 
 Indeed, we now show that the smallest positive eigenvalue
with permutation symmetric eigenfunction 
vanishes in the limit, i.e.  $\lim_{N\to\infty}\lambda_1^{(N)}=0$.  

  Consider
\begin{equation}
\lambda_1^{(N)}
\defeg
\inf_{\psi\in\Sigma_N^+}
\left(\psi,\LCN\psi\right)_{\Lsp^2\left(\Mset^{3N-4}_{\mathbf{u},\vareps}\right)}
\label{LAM_1^N}
\end{equation}
where (using permutation symmetry)
\begin{equation}
\left(\psi,\LCN\psi\right)_{\Lsp^2\left(\Mset^{3N-4}_{\mathbf{u},\vareps}\right)}
=
\frac{N}{2}\int_{\Mset^{3N-4}_{\mathbf{u},\vareps}}\!\!\!\!\!\!
\big(\pdvdue - \pdvuno\big)\psi \cdot 
\frac{\Prm_{\vV_2-\vV_1}^\perp}{|\vV_2-\vV_1|}
\cdot \big(\pdvdue - \pdvuno\big)\psi\,\dd\tau
\end{equation}
Here, $\Sigma_N^+=\{\psi\in\Hsp_s^+\ :\ \norm{\psi}_{\Lsp^2
(\Mset^{3N-4}_{\mathbf{u},\vareps})} =1\}$, and $\Hsp_s^+$ is the permutation
symmetric subspace of $\Hsp^+$.
An upper bound on $\lambda_1^{(N)}$  is obtained by
selecting a suitable trial function $\widehat\psi\in \Sigma_N^+$.
	We use 
\begin{equation}
\widehat\psi
=
A\Big(\sum_{i=1}^N g(\vV_i)-C\Big)
\end{equation}
with $g(\vV)=v_1^2/2$.  In order to satisfy the condition 
$\widehat\psi\in\Sigma_N^+$ one has to choose
\begin{equation}
C=\frac{N}{3}\qquad\qquad A=\frac{3}{2N}
\sqrt{\frac{3N-1}{\abs{\Mset^{3N-4}_{\mathbf{u},\vareps}}}}
\end{equation}
Then, the quantity 
$\left(\widehat\psi,\LCN\widehat\psi\right)_{\Lsp^2
\left(\Mset^{3N-4}_{\mathbf{u},\vareps}\right)}$
can be calculated exactly.  
 In the ``standard" case 
\footnote{
In general, the evolution of $F^{(N)}_t$ on
$\Mset^{3N-4}_{\mathbf{u},\epsilon}$ can always be obtained
from the evolution on $\Mset^{3N-4}_{\mathbf{0},1}$ 
via the simple transformation
$\vVN \to \uVN + \epsilon_0 \vVN$.}
$\mathbf{u}=\mathbf{0}, \vareps=1$, 
a tedious exercise leads to the estimate
\begin{equation}
\lambda_1^{(N)}
<\frac{9}{5\sqrt{\pi}}\frac{1}{\sqrt{3N-4}}
\qquad (N>1)
\end{equation}
Thus, $\lambda_1^{(N)}\to 0$ as $N\to\infty$.  

This is not enough
to conclude rigorously that the BP equation has a vanishing spectral gap
as $N\to\infty$ (because we have not determined the asymptotics for
{\it all} the eigenvalues).  However, in view of our previous 
remarks on the fact that the ellipticity of the equation degenerates
as $N\to\infty$, it seems reasonable to conjecture that this may well
be the case.  In turn, a vanishing spectral gap for the BP equation
would lend support to the conjecture \cite{Vill2} that the Landau
equation itself possesses solutions that decay to equilibrium slower
than exponentially.

 We end on the remark that this result can be easily generalized to
any BP master equation corresponding to a Landau kinetic equation
with ``soft" potentials (for $\gamma <7/3$ in eq.\ (\ref{Qgamma})).  
Conversely, it is not difficult to prove that for 
BP master equations with ``hard" potentials ($\gamma\geq7/3$), 
the spectral gap remains finite in the limit $N\to\infty$.\\[0.5cm] 

\textbf{Acknowledgments}. The work of Lancellotti was supported by NSF 
grant DMS-0207339.
The work of Kiessling was supported by NSF grant DMS-0103808.
We thank Eric Carlen, Maria Carvalho, and Michael Loss for 
helpful conversations about their work. 

\section{Appendix}

 We here recall some basic facts about the Laplacian on a Riemannian manifold. 
Let $\Mset$ be a $n$-dimensional Riemannian manifold with metric $g_{ij}$, 
and let $f$ be a function on $\Mset$, $f\in{\cal C}^\infty(\Mset)$.  
The familiar expression for the Laplace-Beltrami operator acting on $f$ in 
local coordinates is
\begin{equation}
\Delta f
=
g^{-1/2}\,\partial_{x_i}
\left(g^{1/2}\,g^{ij}\,\partial_{x_j}f\right)
\label{LapBel1}
\end{equation}
where $g\equiv{\rm det}(g_{ij})$.  Now, let us take $\Mset$ to be embedded
in the Euclidean space $\Rset^m$, $m>n$, and let $g_{ij}$ be the metric
induced by the standard metric in $\Rset^m$.  Let the function $f$ be defined
and differentiable over some suitable sub-domain of $\Rset^m$ containing $\Mset$.
We associate to any point $x\in\Mset$ a set of orthonormal basis vectors 
$\eV_1,\dots,\eV_n,\eV_{n+1},\dots,\eV_m$ such that $\eV_1,\dots,\eV_n$ 
span the tangent plane $T_x\Mset$, whereas $\eV_{n+1},\dots,\eV_m$ span
the orthogonal complement $T_x\Mset^\perp$ in $\Rset^m$.  
The expression for $\Delta f$ is coordinate-independent, and we are free to
choose local coordinates $(x_1,\dots,x_n)$ such that the coordinate
curves are tangent to the $\eV_1,\dots,\eV_n$.  In such coordinates
$g_{ij}=\delta_{ij}$, $g=1$ and $\partial_{x_j}f=(\nabla f)_j$, i.e
the component 
along $\eV_j$ of the gradient $\nabla f$ of $f$ in $\Rset^m$.  
Then eq.\ (\ref{LapBel1}) becomes
\begin{equation}
\Delta f
=
(P_\Mset\nabla)\cdot[P_\Mset\nabla f]=\nabla\cdot[P_\Mset\nabla f]
\label{LapBel2}
\end{equation}  
where $P_\Mset$ is the orthogonal projection from $\Rset^m$ to $T_x\Mset$.
Note that eq.\ (\ref{LapBel2}) is coordinate-independent in $\Rset^m$,
and preserves permutation symmetry with respect to 
the Cartesian coordinates of $\Rset^m$.  This is important when dealing 
with probability densities for $N$-particle systems, which are constrained 
by the dynamical conservation laws to evolve on certain lower-dimensional 
manifolds and at the same time must be permutation-symmetric.


\end{document}